\documentclass[article,11pt, onecolumn]{IEEEtran}

\usepackage{cite}
\usepackage{hyperref}	
\usepackage{amssymb}
\usepackage{amsmath}
\usepackage{multirow}
\usepackage{cases}
\usepackage{graphicx}
\usepackage{subcaption}
\usepackage{caption}
\usepackage{eurosym}
\usepackage{booktabs,dcolumn}
\usepackage[normalem]{ulem}

\usepackage{color}

\newcommand\mr[1]{\multirow{2}{*}{#1}}
\newcolumntype{d}{D{/}{\hspace{0.1em}/\hspace{0.1em}}{-1}}
\begin{document}

\title{Lighting the Way to a Smart World:\\ Lattice-Based Cryptography for Internet of Things}

\author{Rui Xu, Chi Cheng, Yue Qin, and   Tao Jiang
\IEEEcompsocitemizethanks{\IEEEcompsocthanksitem{ Rui Xu, Chi Cheng, and Yue Qin are with the the Hubei Key Laboratory of Intelligent Geo-Information Processing, School of Computer Science, China University of Geosciences, Wuhan, China.  Tao Jiang  is  with the School of Electronics Information and Communications, Huazhong University of Science and Technology, Wuhan, China. Chi Cheng is the corresponding author. }}
}
\IEEEcompsoctitleabstractindextext{%
\begin{abstract}
The Ukraine power grid cyberattacks remind  us that the smart Internet of Things (IoT) can help us control our light-bulbs, but if under attacks it might also take us into darkness.
Nowadays, many literatures have tried to address the concerns on IoT security, but few of them take into consideration the sever threats to IoT coming from the advances of quantum computing.  As a promising candidate for the future post-quantum cryptography   standard, lattice-based cryptography enjoys the advantages of strong security guarantees and high efficiency, which make  it extremely suitable for IoT applications. In this paper, we summarize the advantages of lattice-based cryptography  and the state of art of their implementations for IoT devices.
\end{abstract}
}

\maketitle

\IEEEdisplaynotcompsoctitleabstractindextext

\IEEEpeerreviewmaketitle

\section{Introduction}
\begin{figure}[!htbp]
\centering
 \includegraphics[width=.7\textwidth]{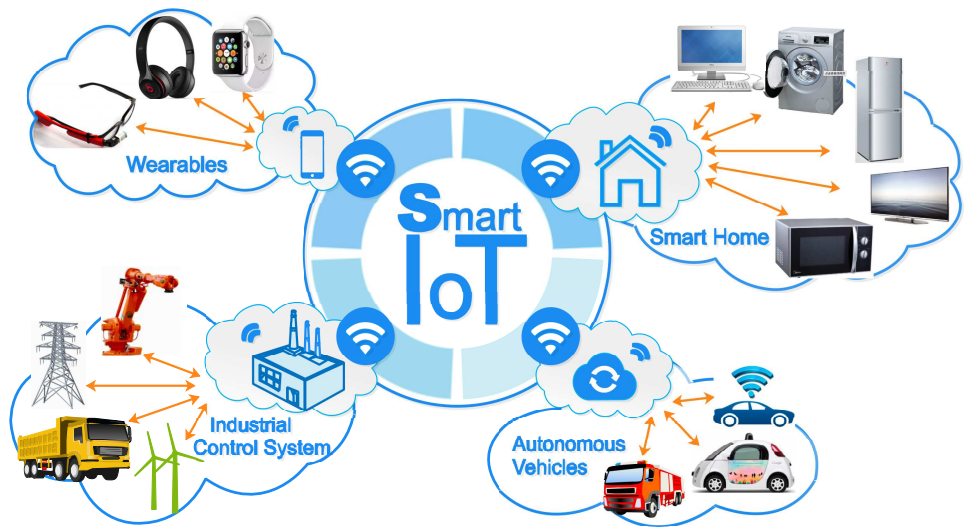}
\caption{Illustration of smart IoT applications.}
 \label{fig1-iot}
\end{figure}

Thanks to the Internet, we are now living in the global village where emails from the U.S. can be transmitted to China within a tenth of second, and real-time teleconferences connect people all over the world. The  Internet of Things (IoT) goes even further beyond, not only affecting the way we exchange data, but also touching the physical world.  Fig. \ref{fig1-iot} shows some scenarios where devices connected to IoT has changed our living:  the smart household appliances in our homes, the wearable gadgets accompany us everyday, the autonomous vehicles, and the industrial control system. In the not too distant future, it would be almost impossible to buy new devices that are not connected to the IoT. And it is estimated that IoT technologies will have an  impact of several trillions to the global economy by 2020 \cite{IEC2016}.

However, the security  and privacy concerns on IoT are always   clouds hanging upon us. As pointed out by Bruce Schneier \cite{Schneier2017}, a security technologist at Harvard University and the chief technology officer of IBM Resilient, the  IoT  companies  are rushing to make their products cheaper and smarter, but without much care about security. The Ukraine power grid cyberattacks remind us that the smart IoT can help us control our light-bulbs, but if under attacks it might also take us into darkness.
Nowadays, many literatures have tried to address the concerns on IoT security  \cite{Fernandes2017}, but few of them take into consideration the sever threat to IoT coming from the advances of quantum computing.

Although quantum computers bear some debates over scientists, with the ever-looming breakthroughs of quantum computing, many researchers are becoming more and more positive about the future of large-scale quantum computers. In March 2017, IBM launched an industry-first initiative, called the IBM Q system, to build a commercially available universal quantum computing system for business and science applications. The publicly available universal quantum processor consists of 15 qubits and their commercially available 17-qubit processor is claimed to be at least twice as powerful.

The quantum threats to cryptography apply equally, or even to a greater extent, to smart objects extensively used in smart IoT services since they involve platforms and systems which are difficult to update. For example, embedded devices in wearables and furnitures are difficult to update  and the scalability issue in IoT devices further complicates the problem. Therefore, we should  taken into consideration post-quantum security   when designing secure architectures and systems for smart IoT, \emph{now}.

Recently, Cheng et al. has called the attention to using post-quantum cryptography (PQC) to secure IoT \cite{Cheng2017}. As a promising candidate for the future PQC standard, lattice-based cryptography enjoys the advantages of strong security guarantees and high efficiency, which make  it extremely suitable for IoT applications. In this paper, we focus on introducing the advantages of lattice-based cryptography  and the state of art of  their implementations for IoT devices.

 In the following, we first give a brief introduction to cryptography and the impact of quantum computers. Then we explain why lattice-based cryptography is a proper choice for smart IoT. Next we give detailed discussions on the state-of-art implementations of lattice-based cryptography on constrained devices, following a high-level overview of lattice based cryptography. Finally we share our opinion on current challenges and directions for future explorations regarding the application of lattice-based cryptography in IoT systems.

\section{Cryptography and Quantum Computers }
Beneath all security protocols, cryptography is used as a fundamental building block.  The canonical implication of security is \emph{confidentiality}, which requires that sensitive information can not be learned by unauthorized party. Symmetric encryption is the simplest and the most popular way of achieving confidentiality. Two communicating parties, Alice and Bob, share a common secret key which is used for both encryption and decryption. Without the knowledge of the secret key, a third party can not learn the encrypted information from the ciphertext.

Symmetric encryption requires a shared common key between two parties, which belongs to the area of symmetric-key cryptography. One drawback of symmetric-key cryptography is the difficulty of establishing secret keys. This is usually done via some costly secure channels such as face-to-face meeting, use of trusted courier or even quantum key distribution. These methods are highly difficult and expensive. Asymmetric-key cryptography (aka public-key cryptography) can be used to overcome this difficulty as it provides a mechanism to distribute cryptographic keys over insecure channel. In public-key cryptography, Alice has a pair of related keys: one is the private key and the other is called the public key.  The private key, as suggested by its name, is kept private to Alice herself while her public key is known to everyone.

Using public-key encryption algorithms, everyone can encrypt message and send it to Alice using her public key. But only Alice who has the private key is able to decrypt. This feature allows Bob to encrypt a secret session key of a symmetric encryption scheme such as AES and transmit it to Alice. After decrypting, Alice gets the key for AES and can now establish a secure channel with Bob via AES using the session key. This is called hybrid encryption and is used in many security protocols such as Transport Layer Security (TLS) protocol. Another method known as the key exchange protocol allows Alice and Bob to negotiate session key over an  insecure channel.

Yet another problem arises. How can Bob, or anyone, make sure that the claimed public key for Alice indeed belongs to Alice but not Eve? This involves the notion of \emph{trust} in cryptography. Generally two solutions are available. One is to use the Public Key Infrastructure (PKI) and the other is to use Identity Based Encryption (IBE).

PKI is a mechanism that can bind the public keys with the identities of their owners. A trusted certificate authority (CA) can issue certificate to an entity to prove that the public key indeed belongs to this entity. Informally a certificate can be viewed as a digital signature made by the CA (using its private key) on the message that ``This public key belongs to Alice''. A digital signature of a message is a digital counterpart to the hand-written signature which assures that the message is generated by the signer (this relates to \emph{authentication} in cryptography). Everyone can use the CA's public key to verify the validity of the CA's signature so as to verify the certificate. Of course as CA is trusted, its public key must be well known. This can be easily achieved since trusted CAs (like government agencies or global organizations) usually have large influence and rich resources to distribute their public keys to the public.

The other method of using IBE also requires a trusted authority to generate the public and private key pair of an entity. But no certificate is needed. In an IBE system, the public key of an entity can be anything so an entity can use its identity, such as name of an organization, email address of a person,  as its public key which can be easily verified by others. The PKI mechanism requires users to verify each certificate issued by CAs. Thus heavy public-key operations are needed in PKI, which are obviously not friendly to IoT applications. IBE can efficiently reduce the cost to verify the correctness of public keys, which turns to be favorable in the scenario of IoT.

Modern cryptography bases its security on rigorous proofs for assuring security in extreme adversarial situations. The acknowledged security of essentially all provably secure cryptographic primitives is reduced to the confidence on well-established hardness of some  mathematical problems.  The integer factorization problem and (Elliptic Curve) discrete logarithm  problem are two famous problems of this kind. They are the bases for RSA, Diffie-Hellman and Elliptic Curve Cryptography (ECC), which are widely used in today's cryptography. The best known classical algorithms (on Turing machines) for solving factorization and discrete logarithm problem work with sub-exponential time complexity. But Shor's quantum algorithm can solve both within polynomial time. A direct consequence is that once large-scale quantum computers are available our current public-key cryptography system such as RSA and ECC, would be \emph{completely broken}. Hence, it is of high priority that we explore alternative problems which are intractable for both classical computers and quantum computers.

Another mild yet universally influential impact of quantum computing techniques comes from Grover's algorithm which presents a quadratic speedup for searching problems over classical algorithms. Grover's algorithm can be used for many cryptanalysis methods which require some sort of brute force. For example guessing the secret key of AES can be accelerated using Grover's algorithm. Generally speaking, one can simply double the length of the key to achieve the same post-quantum security level regarding the impact of Grover's algorithm.

The  quantum threat has been well recognized by government agencies, large corporations and academic researchers all over the world. The alternative solution called PQC, which aims to provide cryptographic solutions those remain secure even the adversary has access to large-scale quantum computers, is now a hot and steadily growing topic. National Security Agency (NSA) announced, in 2015, their preliminary plans for transitioning to quantum resistant cryptography for protecting classified information. In December 2016, National Institute of Standards and Technology (NIST) issued an open call for standardization consideration of post quantum cryptographic algorithms. At the time of writing (December 2017), the open call is finished. NIST is arranging the first PQC standardization conference, to be held in April, 2018, for the submitters to present and discuss their submissions. Currently Google is experimenting post-quantum cryptography in its web browser Chrome. The Tor (a software which protects its users against Internet surveillance) project is also trying to implement lattice-based key exchange protocols to achieve post-quantum security.


\section{Why Lattice-Based Cryptography?}
Different proposals have been proposed to achieve post-quantum security including hash-based signatures, code-based cryptography, multi-variate polynomial-based cryptography, and lattice-based cryptography. We focus on lattice-based cryptography in this article. In our opinion, lattice-based cryptography is highly suitable for smart IoT applications. Firstly, the strong security guarantees and high efficiency shown by lattice-based cryptography make it extremely suitable for IoT applications. Secondly, the wide applicability of lattice-based cryptography can accommodate further advances of smart IoT services. Last but not least, lattice-based cryptography receives the most intensive attention among all subfields of post-quantum cryptography. The recent NIST call has received 82 submissions for post-quantum cryptographic algorithms and 28 of them are based on lattice, taking the lead.

Lattice-based cryptography has strong security guarantees. The underlying hard problems  are intensively studied for decades but no efficient algorithm, both classically and quantumly, is known for those problems. Moreover, lattice-based cryptography enjoys worst-case to average-case reduction. Cryptography inherently requires average-case intractability considering the  requirement of random keys. The worst-case to average-case reduction essentially guarantees that lattice-based cryptography is secure on average unless every instance of the underlying lattice problem is easy. From the practical aspect, this worst-case reduction makes parameter selection and key generation much easier in lattice-based cryptography. For example, the RSA cryptosystem is based on the hardness of integer factorization. But this is an worst-case problem. It is known that if the primes have certain number-theoretic properties, the problem turns out to be essentially easy. Hence it is important to \emph{avoid such structures} in key generation for RSA. Unfortunately, we do not know whether such structures have been fully explored. In contrast, lattice-based cryptography is based on average-case hard problems. When generating keys for lattice-based cryptography, one only needs to select proper \emph{parameter size} and then generate keys uniformly.

Lattice-based cryptographic algorithms operate over relatively smaller integers, compared with large integers  used in RSA. The computations involved in the state of art of lattice-based algorithms mainly consists of simple operations between matrices and vectors in some rings or fileds of small order.    Actually lattice-based cryptography runs faster than RSA   and it can be implemented on low-power devices with 8-bit microcontrollers. Recent implementations of lattice-based cryptography have been already  an order of magnitude faster than the corresponding
RSA implementations. For example, the current state-of-art implementation of R-LWE based encryption on 8-bit AVR microconstroller can finish an encryption within 2 million cycles, while the RSA-1024 (has a lower level of security and no post-quantum security) implementations on comparable devices need more than 23 million cycles for the same task \cite{Liu2017}.

Other candidates of post-quantum cryptography, for example the code-based cryptography, may present even better performance regarding computational efficiency but inevitably require larger sizes for keys and ciphertexts. We stress that it is the balance  among performance metrics, such as key size, ciphertext and signature lengths, computational efficiency and confidence of security, that make lattice-based cryptography a well fit for IoT applications.

\begin{figure}
\centering
        \includegraphics[width=.6\textwidth]{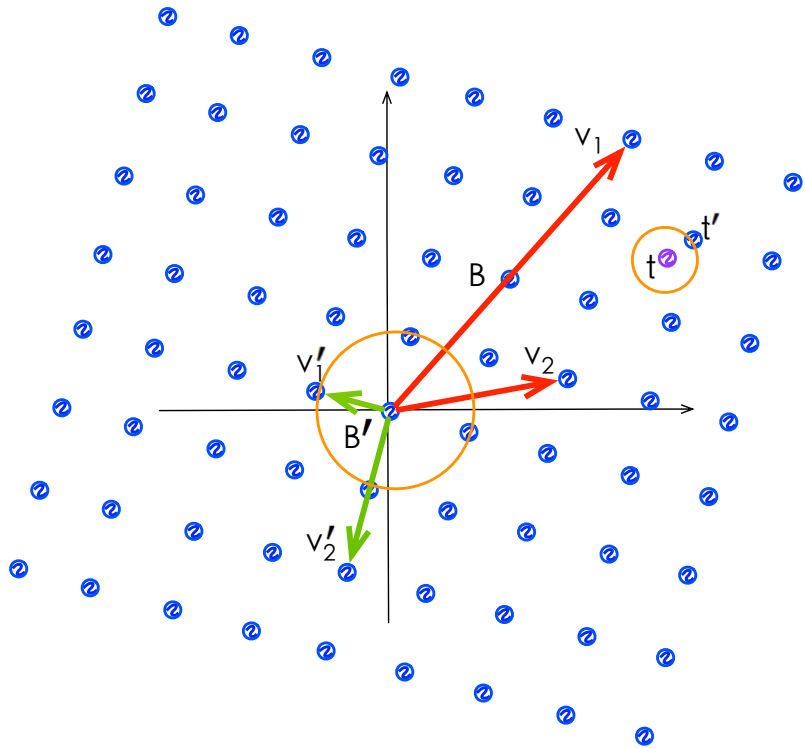}
\caption{Illustration of a 2-Dimensional lattice}
\label{fig2-svp}
 \end{figure}

\section{An Introduction to Lattice Based Cryptography}
Since the impact of quantum algorithms is mild for symmetric-key cryptography but devastating for current public-key cryptography, many efforts have been put forth in the research of PQC to seek for replacement of universally used cryptosystems such as RSA and ECC. Among them, lattice-based cryptography is a very promising candidate. See Peikert \cite{Peikert2016} for a comprehensive survey on lattice-based cryptography.

Lattice-based cryptography is based on the   hardness of solving some   geometric problems over high-dimensional lattices, such as the shortest vector problem (SVP) and the closest vector problem (CVP). But these problems are easy to solve if one has a good basis. A good basis consists of short vectors which are nearly orthogonal, while a bad basis consists of long vectors which generally point in the same direction. In Fig. \ref{fig2-svp}, the points  coloured in blue  are lattice vectors which are integer  combinations of the basis  $\mathbf{B} = \{\mathbf{v}_1, \mathbf{v}_2\}$. The  basis $\mathbf{B}$, colored in red, is a bad basis and the  basis $\mathbf{B}'$, colored in green, is a good basis consisting of almost orthogonal vectors. The SVP   is to find a shortest nonzero vector such as the vector $\mathbf{v}'_1$. For an arbitrary vector in the space such as the purple point $\mathbf{t}$, the CVP is to find a lattice vector $\mathbf{t}'$ which is the closest to $\mathbf{t}$. Note that we use 2-dimensional (2D) examples for easy visualization but SVP and CVP are easy for 2D case. They get (exponentially) harder as the dimension of lattice grows.

\begin{figure}
\centering
        \includegraphics[width=.6\textwidth]{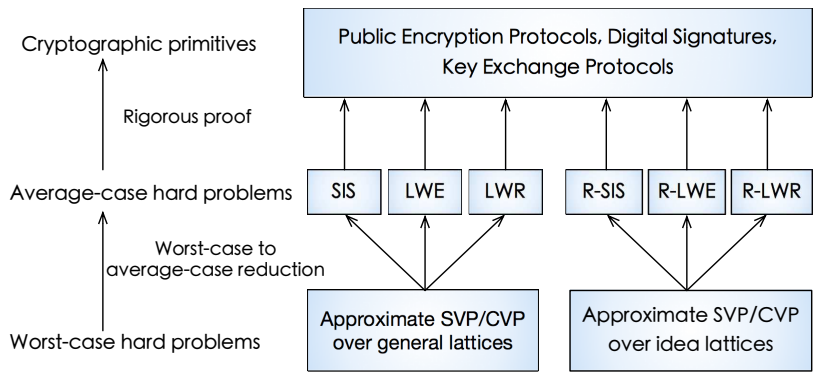}
\caption{Landscape of hard problems in lattice-based cryptography. }
\label{fig3-lattice}
 \end{figure}

In Fig. \ref{fig3-lattice}, we give a landscape of the hard problems in lattice-based cryptography. The above mentioned SVP and CVP are called \emph{worst-case} hard problems. This type of problems only guarantee that there exist instances that are intractable. But the problem might be easy on average. One big advantage of lattice-based cryptography is that there are many \emph{average-case} problems such as the short integer solution (SIS) and learning with errors (LWE) problems. These are well encapsulated average problems enjoying a worst-case to average-case reduction which states that SIS and LWE  are hard on average (for a random instance) unless the related problems on lattices are easy for \emph{all} instances.  The worst-case to average-case reduction gives the lattice-based cryptography a easy way to  construct  cryptographic schemes and prove their security. That is, one can work on the conceptually simple average-case problems to construct cryptographic primitives while at the same time gains confidence about the constructions from the low-level hard lattice problems in the worst case.

The SIS and LWE problems can be easily described in the form of solving linear equation systems. The SIS  problem is to find short nontrivial (excluding the all-zero solutions) integer solutions to an homogeneous linear system  with the coefficient matrix uniformly randomly generated. The   LWE problem asks to  find the  the secret vector $\textbf{s}$, given polynomially many samples $A\textbf{s}+\textbf{e}$, where $A$ is a uniformly generated matrix  and    $\textbf{e} $ represents the noises  selected from some specified error distribution.  More compact ring versions of SIS and LWE are called Ring-SIS (R-SIS) and Ring-LWE (R-LWE).  These ring variants can  reduce memory requirement and computational costs for the cryptographic schemes.
 
In practice, besides the primitives based on (R-)SIS and (R-)LWE, NTRUEncrypt is a lattice-based encryption scheme standardized by the  industry. The security of NTRUEncrypt is based on  hard problems over the so called NTRU lattices. Although there is no worst-case reduction for standard NTRUEncrypt, it has withstood attacks for 20 years (with update in the parameters). Another problem called learning with rounding (LWR) is a derandomization of LWE, where no error distribution is needed  to boost the  performance of related schemes.

\section{Implementations of Lattice-Based Cryptography for Resource-Constrained Devices}
\label{section-recomm}
The ambition of connecting everything inherently raises the challenge of implementing cryptographic algorithms on resource-constrained devices such as sensors and actuators. In this section we review the state of art of implementations of lattice-based cryptography on constrained devices, which can be divided into software implementations on microcontrollers and hardware implementations on FPGAs.

Main operations of lattice-based cryptographic constructions involve matrix-vector multiplication (schemes based on SIS and LWE) and polynomial multiplications (schemes based on R-SIS, R-LWE, and NTRU).   Polynomial multiplication used in R-LWE based schemes can be optimized via the number theoretic transform (NTT) method, which is a  discrete Fourier transform (DFT) variant. In NTT, the primitive root  of unity modulo an integer $N$ is used instead of complex roots of unity used in DFT. The NTT can transform the classic polynomial multiplication to point-wise multiplication to reduce the complexity from quadratic to quasi-linear. The underlying algorithm can be adjusted to reduce the number of necessary NTT transformations as suggested by Roy et al. for the R-LWE based encryption \cite{Roy2014}. Optimization methods for FFT can also be borrowed.

Many lattice-based cryptographic constructions (LWE and R-LWE based) require sampling from discrete Gaussian distribution \cite{Howe2016}. There are many methods to implement the discrete Gaussian sampler including rejection sampling, cumulative distribution table (CDT) sampling, Bernoulli sampling and Knuth-Yao sampling. Sampling from a discrete Gaussian is difficult due to the fact that it requires high precision computation of the exponential function or large precomputed tables.


In the following we discuss implementations of  lattice-based proposals for public-key encryption, digital signatures and key exchange protocols. We summarize the  implementations  on low-cost microcontrollers in Table \ref{tab1}, in which the `$/$' within a cell is used to separate figures for two different operations in the underlying algorithm. The two operations are encryption and decryption for public-key encryption schemes; signing and verification for signature schemes; sever-side computation and client-side computation for key exchange protocols. The `ROM' presents the memory used by the implementation. The hardware implementations of lattice-based cryptography on FPGAs is summarized  in Table \ref{tab2}.
\begin{table}[!htb]
\centering
\caption{Software implementations of lattice-based cryptography on low-cost microcontrollers.}
\label{tab1}
\begin{tabular}{|c|c|c|c|c|c|c|c|}
\hline
  \mr{Schemes} &  \mr{Bit Security} & \multicolumn{3}{c|}{Platform} & \mr{Cycles} & \mr{Time   (ms)} & \mr{ROM (KB)}  \\ \cline{3-5}
  &    & Device & CPU & MHz &   &    & \\
  \hline
  NTRUEncrypt \cite{Guillen2017}                  & 128 (pre)                                                      & \begin{tabular}[c]{@{}c@{}}Cortex-M0\\ (XMC1100)\end{tabular}       & 32-bit & 32  &   588,044 / 950,371      &  18.4 / 29.7                 & 9                         \\ \hline
  R-LWEenc \cite{Liu2017}                     & \begin{tabular}[c]{@{}c@{}}106 (pre) \\ 46 (post)\end{tabular} & ATxmega128                                                           & 8-bit  & 32  &    796,872 / 215,031   &  24.9 / 6.7            & 6.5                       \\ \hline
  \multirow{2}{*}{R-BIN-LWEenc \cite{Buchmann2016}} & \multirow{2}{*}{94 (pre)}                                      & ATxmega128                                                          & 8-bit  & 32  & 1,573,000 / 740,000      &  49.2 / 23.1                  & 2.7                       \\ \cline{3-8}
                                &                                                                & Cortex-M0                                                           & 32-bit & 32  & 999,000 / 437,000         &  31.2 / 13.7                & 5.6                       \\ \hline
  \multirow{2}{*}{IBE \cite{Guneysu2017}}          & \multirow{2}{*}{80 (pre)}                                      & Cortex-M0                                                           & 32-bit & 32  &  3,297,380 / 1,155,000    &  103.0 / 36.1               & 17                        \\ \cline{3-8}
                                &                                                                & Cortex-M4                                                           & 32-bit & 168 & 972,744 / 318,539        &  5.8 / 1.9                  & 18.7                      \\ \hline
  BLISS \cite{Liu2017}                         & 128 (pre)                                                      & ATxmega128                                                          & 8-bit  & 32  &  10,156,247 / 2,760,244    &  317.4 / 86.3               & 18.4                      \\ \hline
  \multirow{2}{*}{NewHope \cite{Alkim2016}}      & \multirow{2}{*}{128 (post)}                                    & \begin{tabular}[c]{@{}c@{}}Cortex-M0\\ (STM32F051R8T6)\end{tabular} & 32-bit & 48  & 1,467,101 / 1,738,922   & 30.6 / 54.3             & 30.2                      \\ \cline{3-8}
                                &                                                                & \begin{tabular}[c]{@{}c@{}}Cortex-M4\\ (STM32F407VGT6)\end{tabular} & 32-bit & 168 &  860,388 / 984,761       &  5.1 / 5.9                  & 22.8                      \\ \hline
  \end{tabular}
  \end{table}

As the first lattice-based encryption scheme, the NTRUEncrypt encryption has been accepted by the IEEE P1363.1 standard. The NTRU cryptosystem was patented by its inventors along with a variant using `product-from keys' for efficient implementation. But recently (March, 2017) they announced their decision for placing all of its NTRUEncrypt patents in the public domain. Recently Guillen et al. \cite{Guillen2017} explored the feasibility of employing NTRUEncrypt in constrained devices (a Cortex-M0 based microcontroller).

 The R-LWE based encryption scheme is similar to NTRUEncrypt regarding communication and computational costs. One advantage of R-LWE based encryption is its provable security but it comes at the price of a high-precision Gaussian sampler. 	Liu et al. \cite{Liu2017} presented a constant-time implementation of the R-LWE based encryption scheme with 46-bit post-quantum security level on an 8-bit ATxmega128 microcontroller (32 MHz, 128 KB flesh memory, 8 KB RAM) with encryption and decryption time of 24.9 ms and 6.7 ms, respectively. Buchmann et al. \cite{Buchmann2016} proposed an encryption scheme by replacing the Gaussian noise in R-LWE with a binary distribution and implemented the scheme (R-BIN-LWEenc) on low-cost microcontrollers.

\begin{table}[!htb]
\centering
\caption{Hardware implementations of lattice-based cryptography on FPGAs. }
\label{tab2}
\begin{tabular}{  |c|c|c|c|c|c|}
\hline
\multirow{2}{*}{Schemes}                           & \multirow{2}{*}{Bit Security} & \multirow{2}{*}{Devices} & \multirow{2}{*}{MHz} & \multirow{2}{*}{Cycles} & \multirow{2}{*}{Time($\mu$s)} \\
   & & & & & \\  \hline
R-LWEenc \cite{Roy2014}                         & 128 (pre)                    & V6LX75T                  & 313                 & 6,300 / 2,800             & 20.1 / 9.1                       \\ \hline
\multirow{2}{*}{R-LWEenc \cite{Poppelmann2014}} & \multirow{2}{*}{128 (pre)}    & \multirow{2}{*}{S6LX9}  & 144                    & 136,986                 & 946                            \\ \cline{4-6}
                                                  &                               &                         & 189                    & 66,338                  & 351                            \\ \hline
\multirow{2}{*}{BLISS \cite{Poppelmann2014b}}   & \multirow{2}{*}{128 (pre)}    & \multirow{2}{*}{S6LX25} & 129                & 16,210                  & 126.6                          \\ \cline{4-6}
                                                  &                               &                         & 142                & 9,835                   & 69.3                           \\ \hline
IBE \cite{Guneysu2017}                          & 80 (pre)                      & S6LX25                  & 174                     & 13,958 / 9,530            & 80.2 / 54.8                      \\ \hline
\end{tabular}
\end{table}

 G\"{u}neysu and Oder \cite{Guneysu2017} demonstrated that IBE has become practical even for embedded devices such as Cortex-M microcontrollers and FPGAs by implementing a R-LWE based IBE scheme. Many FPGA implementations of R-LWEenc exist, we introduce two of them. Roy et al. \cite{Roy2014} presented an FPGA implementation of R-LWEenc optimized for throughput on V6LX75T. P\"{o}ppelmann and G\"{uneysu} \cite{Poppelmann2014} presented an area-optimized implementation of R-LWEenc on S6LX9 by carefully choosing parameters to eliminate modular reduction operations.

We refer the readers to Howe et al. \cite{Howe2015} for a comprehensive discussion on lattice-based signatures. Original proposals for lattice-based signature schemes such as the GGH signature and NTRUSign, which utilize the hardness of CVP, have been broken. The current state-of-art signature scheme  is BLISS, which is based on R-LWE and has been proven secure in the random oracle model. Discrete Gaussian sampling accounts for more budget in signature schemes than that in encryption schemes because the deviation used in signature schemes is much larger.   State-of-art software implementation and hardware implementation of BLISS are Liu et al. \cite{Liu2017} and P\"{o}ppelmann et al. \cite{Poppelmann2014b}.  Ducas proposed a variant called BLISS-B which can reduce the repetition rate and in turn speeding up the key generation by a factor of 5 to 10.

    NewHope is a post-quantum key exchange protocol based on R-LWE and has been used by Google's post-quantum security experiments within Chrome. Alkim et al. \cite{Alkim2016} presented a software implementation of NewHope for Cortex-M family of 32-bit microcontrollers. With various generic and platform-specific optimizations, their implementation demonstrated that lattice-based key exchange protocols are indeed promising candidates for post-quantum IoT security. The Cortex-M0 implementation requires about 1.5 million cycles for server side computation and 1.8 million for client side. On the more powerful M4 platform the corresponding cycles are 0.8 million and 9.8 million.


 \section{Challenges and Future Research Directions}
Various implementation results have demonstrated that lattice-based cryptography is practical even for resource-constrained devices. Regarding computational speed, lattice-based cryptography is already faster than traditional public-key cryptography such as RSA or even ECC. But one can not draw the conclusion that the former performs better in practice since lattice based cryptography usually requires more communication cost which is much more resource-consuming. Further improvement of lattice-based cryptography remains a challenge. The implementation optimization is of course necessary, but theoretical improvement for reducing ciphertext and signature size might be more promising. Further directions include tighter security proof, efficient construction, reducing the use of discrete Gaussian noise and efficient encoding techniques.

Most of implementations we discussed do not provide protection against side-channel attacks (SCAs). It is of prominent significance to provide side-channel-attack-resistant implementations for smart IoT applications since they are more venerable to SCAs.

The provable security of most lattice-based cryptography does not guarantee security in practice and might even cause overlook on practical security. Choosing appropriate parameters for lattice-based schemes is another challenge. Many incomparable algorithms for analyzing lattice problems are available and the performance for some of them are not well understood. A unified model for evaluating security level of lattice-based cryptography is highly desirable. In R-LWE based constructions, the parameters are also constrained by the requirement of NTT-friendly choices which may lead to gaps in security level.

Analyzing the security of lattice-based cryptography against fully quantum attacks is more than a theoretical interest. Current PQC only considers the impact of quantum computers (or quantum algorithms). However, in the quantum world the attacker is able to have quantum interaction with the cryptosystem. The quantum random oracle model in the literature is one kind of this directions.


\section*{Acknowledgments}
 The work presented in this paper was supported in part by the National Natural Science Foundation of China under Grant no. 61672029.




\end{document}